\documentclass[preprint,12pt]{aastex}

\usepackage{emulateapj5}

\slugcomment{Accepted to ApJ}

\shorttitle{L Subdwarf 2MASS 0532+82}
\shortauthors{Burgasser et al.}

\begin{document}

\title{The First Substellar Subdwarf?  Discovery of a Metal-poor L Dwarf with Halo
Kinematics}

\author{
Adam J.\ Burgasser\altaffilmark{1,2},
J.\ Davy Kirkpatrick\altaffilmark{3},
Adam Burrows\altaffilmark{4},
James Liebert\altaffilmark{4},
I.\ Neill Reid\altaffilmark{5},
John E.\ Gizis\altaffilmark{6},
Mark R.\ McGovern\altaffilmark{1},
L.\ Prato\altaffilmark{1},
\&
Ian S.\ McLean\altaffilmark{1}
}

\altaffiltext{1}{Department of Physics \& Astronomy,
University of California
at Los Angeles, Los Angeles,
CA, 90095-1562; adam@astro.ucla.edu, mcgovern@astro.ucla.edu,
lprato@astro.ucla.edu, mclean@astro.ucla.edu}
\altaffiltext{2}{Hubble Fellow}
\altaffiltext{3}{Infrared Processing and Analysis Center, M/S 100-22,
California Institute of Technology, Pasadena, CA 91125; davy@ipac.caltech.edu}
\altaffiltext{4}{Steward Observatory, University of Arizona,
Tucson, AZ 85721; burrows@as.arizona.edu, liebert@as.arizona.edu}
\altaffiltext{5}{Space Telescope Science Institute, 3700 San
Martin Drive, Baltimore, MD 21218; inr@stsci.edu}
\altaffiltext{6}{Department of Physics and Astronomy, University
of Delaware, Newark, DE 19716; gizis@udel.edu}

\begin{abstract}
We present the discovery of the first L-type subdwarf, 2MASS J05325346+8246465.
This object exhibits enhanced collision-induced H$_2$ absorption, resulting in
blue NIR colors ($J-K_s = 0.26{\pm}0.16$).
In addition, strong hydride bands
in the red optical and NIR, weak TiO absorption,
and an optical/J-band spectral morphology similar
to the L7 DENIS 0205$-$1159AB imply a
cool, metal-deficient atmosphere. We find that 2MASS 0532+8246 has both a
high proper motion, $\mu$ = 2$\farcs$60$\pm$0$\farcs$15 yr$^{-1}$, and a
substantial radial velocity, $v_{rad} = -195{\pm}11$ km s$^{-1}$,
and its probable proximity to the Sun (d = 10--30 pc)
is consistent with halo membership.
Comparison to subsolar-metallicity
evolutionary models strongly
suggests that 2MASS 0532+8246 is substellar, with a mass of
0.077 $\lesssim$ M $\lesssim$ 0.085 M$_{\sun}$ for ages
10--15 Gyr and metallicities $Z = 0.1-0.01$ $Z_{\sun}$.
The discovery of this object clearly indicates that star formation occurred
below the Hydrogen burning mass limit at early times, consistent with prior results
indicating a flat or slightly rising mass function for the lowest-mass
stellar subdwarfs.
Furthermore, 2MASS 0532+8246 serves as a prototype for a new spectral class
of subdwarfs, additional examples of
which could be found in NIR proper motion surveys.
\end{abstract}


\section{Introduction}

Subdwarfs are metal-deficient stars, classically defined as lying below
the stellar main-sequence in optical color-magnitude diagrams \citep{kui39}.
These objects are in fact not subluminous but rather hotter (i.e., bluer in
optical colors) than equivalent mass
main-sequence dwarfs, a consequence of their reduced metal opacity
\citep{cha51,san59}.  Cool subdwarfs (spectral types sdK and sdM) are typically
found to have halo kinematics (${\langle}V{\rangle} = -202$ km s$^{-1}$; Gizis 1997),
and these objects are presumably relics of the early
Galaxy, with ages of 10--15 Gyr.  Because low mass subdwarfs have lifetimes far in excess of
the age of the Galaxy, they are important tracers of Galactic chemical history,
and are representatives of the first generations of star formation.

All of the coolest subdwarfs ([Fe/H] $\sim -1.2\pm$0.3) and extreme subdwarfs
([Fe/H] $\sim -2.0\pm$0.5; Gizis 1997) currently known
have been identified in optical
proper motion surveys, most notably Luyten's Half-Second Catalog \citep[LHS]{luy79a}
and Two-Tenths Catalog \citep{luy79b}, the APM Proper
Motion Survey \citep{shz00}, and the Galactic plane survey of \citet{lep02}.
The substantial space velocities of halo subdwarfs allow them to stand out
in these surveys amongst the overwhelming multitude of similarly faint,
but slowly moving, background stars.
However, only a handful of very cool subdwarfs with spectral
types sdM6/esdM6 and
later\footnote{Based on the Gizis (1997) classification scheme.}
have been found \citep{giz97,giz97b,sch99,lep03a,lep03b}.
These stars exhibit characteristic spectral
signatures of strong metal hydrides (CaH, MgH, AlH, FeH), weak and/or absent
metal oxides (TiO, VO, CO), and enhanced collision-induced (CIA) H$_2$
absorption \citep{mou76,lie87,sau94,leg00}.  With T$_{eff} \gtrsim$ 2900 K \citep{leg00},
the coolest subdwarfs known have masses
just above the Hydrogen Burning Minimum Mass
(HBMM)\footnote{The HBMM is defined here as the
minimum mass for which core Hydrogen fusion
balances luminosity at ages later than 10 Gyr.}, which ranges from
0.072 M$_{\sun}$ for Solar composition ($Z = Z_{\sun}$)
to 0.092 M$_{\sun}$ for $Z = 0$ \citep{cha97,bur01}.

In contrast, hundreds of significantly cooler Population I, or disk,
dwarfs have been identified\footnote{A listing of known M, L, and T dwarfs is maintained
by J.\ D.\ Kirkpatrick at \url{http://spider.ipac.caltech.edu/staff/davy/ARCHIVE/index.html}.},
predominately in the
wide-field optical and near-infrared (NIR)
surveys DENIS \citep{epc97}, 2MASS \citep{skr97}, and SDSS \citep{yor00}.
These discoveries
include many dozens of substellar objects, and extend well
beyond the M spectral class into
the L \citep{kir99,mrt99} and the
T \citep{me02a,geb02} classes.  The success of recent surveys in identifying
very cool dwarf star and brown dwarfs derives from their use of red-sensitive CCDs
and infrared array detectors.
As we proceed to cooler temperatures, the majority of emitted flux from
both disk dwarfs and subdwarfs moves to NIR wavelengths.
Unfortunately, proper
motion surveys of sufficient temporal breadth to identify
halo stars rely
primarily on optical, typically photographic, imaging.
Factoring in their intrinsic rarity (0.3\% number density compared to
disk dwarfs; Reid \& Hawley 2000) and substantial evolution
toward lower luminosities \citep[see Figure 3]{bur93,cha97}, it is clear why few
very cool subdwarfs have been found.

In this article, we present the discovery of a new cool subdwarf, identified in the
2MASS database.  This object,
2MASS J05325346+8246465\footnote{Source designations for the 2MASS Point Source Catalog are
given as ``2MASS Jhhmmss[.]ss$\pm$ddmmss[.]s''.  The suffix conforms
to IAU nomenclature convention and is the sexigesimal R.\ A.\ and
decl.\ at J2000 equinox.  We adopt a shorthand notation of ``2MASS hhmm$\pm$ddhh'' throughout
the remainder of this Letter.}, is a high motion source with a
spectral morphology consistent with a metal deficient L dwarf.
In $\S$ 2 we discuss the identification of 2MASS 0532+8246 and subsequent NIR and optical
observations obtained at the Palomar 60'' Telescope.  In $\S$ 3 we describe spectroscopic
observations of 2MASS 0532+8246 obtained with the Near-IR Spectrometer
\citep[hereafter NIRSPEC]{mcl98,mcl00} and Low
Resolution Imaging Spectrograph \citep[hereafter LRIS]{oke95} at Keck Observatory.
In $\S$ 4 we derive the space motion of this object from the imaging and spectroscopic data.
In $\S$ 5 we discuss the substellarity of 2MASS 0532+8246 based on subsolar metallicity models
from \citet{bur01}.  We discuss our results in $\S$ 6.

\section{Identification of 2MASS 0532+8246}

2MASS 0532+8246 was identified in a search for T dwarfs using the 2MASS Working
database \citep{me02a}.  It was selected for its blue NIR colors
($J-K_s = 0.26{\pm}0.16$) and lack of an optical counterpart at the 2MASS position
in both POSS-I and POSS-II photographic plates.
A finder chart is given in Figure 1, and astrometric and photometric
data from 2MASS are listed in Table 1.
Follow-up NIR
imaging using the Palomar 60''
Infrared Camera \citep[hereafter IRCam]{mur95} on 25 September 1999 (UT)
determined that this object is not
an uncatalogued minor planet.
Gunn $r$- and $z$-band images obtained with the Palomar 60'' Facility CCD
Camera on 18 November 2002 and 25 January 2003 (UT) further
confirmed the red optical/NIR colors of this object.
Aperture photometry from 18 November 2002, calibrated with
observations of the SDSS photometric standard
BD +33$\degr$ 4737 \citep[correcting ($r^{\prime}, z^{\prime}$) $\rightarrow$
(gunn $r$, gunn $z$) using photometry from Kent 1985]{smi02}
yield $r > 20.9$ (5$\sigma$ upper limit) and $z = 18.30{\pm}0.16$.
As the NIR and optical/NIR colors of this object are
consistent with those of other identified late-type L and
T dwarfs \citep{kir99,fan00,dah02}, 2MASS 0532+8246 was initially
believed to be a firm T dwarf candidate.

\section{Spectroscopic Observations}

\subsection{NIR Spectrum}

We obtained NIR spectra of 2MASS 0532+8246 on 24 December 2002 (UT)
using NIRSPEC on the Keck-II 10m Telescope.
Conditions were clear with seeing of 0$\farcs$6.
We observed the object in three spectral configurations,
N3 (1.14 -- 1.38 $\micron$), N5 (1.51 -- 1.79 $\micron$), and
N6b (1.94 -- 2.32 $\micron$), corresponding roughly to the J, H, and K NIR
bands.  Spectral resolution was R $\approx$ 2000 for our 0$\farcs$38 slit.
For each observation, two exposures of 300s each were obtained,
dithering 20$\arcsec$ along the slit.  All observations were made in the airmass
range 2.6--3.0.  Immediately after each sequence of exposures,
internal flat field and NeAr arc lamps were observed for calibration, and
the A0 star HD 34360 was observed just before
or after the target source in each spectral order at similar airmasses.
Data were reduced using the REDSPEC
package\footnote{See \url{http://www2.keck.hawaii.edu/inst/nirspec/redspec/index.html}.};
a detailed description of our reduction procedures is given
in \citet{mcl03}.
Final flux calibration of the individual spectral orders was done using 2MASS photometry;
however, because the orders do not span the entire photometric bandpasses \citep{cut03},
our final flux calibration may be somewhat underestimated
at J-band, but
reasonably correct for the H- and K-bands.

The resulting
NIR spectrum is shown in Figure 2a.  The 1.5--2.4 $\micron$ spectral energy distribution
is fairly smooth and blue, reminiscent of an early-type main-sequence star or white dwarf,
and quite unlike the much redder NIR spectra of late-type M and L dwarfs (cf.\ overlay of
the L7 DENIS 0205$-$1159AB in Figure 2a;
Delfosse et al.\ 1997; Kirkpatrick et al.\ 1999; McLean et al.\ 2003).
However, the red optical/NIR colors and
optical molecular bands (see below) of 2MASS 0532+8246 are inconsistent with a hot stellar atmosphere.
We instead attribute the shape of the NIR spectrum to enhanced absorption by
the CIA H$_2$ 1-0 quadrupole band centered
near 2.5 $\micron$ \citep{sau94,bor97}.  This pressure-sensitive band, present in
late-type L and T dwarfs \citep{tok99,me02a} and cool subdwarfs \citep{leg00},
is generally broad and featureless,
consistent with the observed spectrum.  Strong H$_2$ absorption may explain the
absence of CH$_4$ absorption bands at 1.6 and 2.2 $\micron$, the defining features of T dwarfs;
and the 2.3 $\micron$ CO band,
a key spectral feature in M, L, and early-type T dwarf spectra.
On the other hand, H$_2$O absorption appears to be
present at 1.3--1.5 and 1.7--2.0 $\micron$, and minor features throughout the H-band are likely
attributable to weaker H$_2$O lines.
Two weak features at 1.573 and 1.626 $\micron$ are present,
possibly attributable to FeH absorption
\citep{cus03} or poor correction of telluric OH lines;
these features are not coincident
with the 1.6 $\micron$ CH$_4$ band.
Overall, this portion of the spectrum of 2MASS 0532+8246 is quite unlike any known
late-type M, L, or T dwarf, based on its blue slope
and absence of CH$_4$ and CO bands.

The J-band spectrum,
shown in detail in Figure 2b, is far more rich, with
atomic lines of K I (1.1690, 1.1773, 1.2543, 1.2522 $\micron$)
and Fe I (1.1528, 1.1886, 1.1976, 1.2128, 1.3210 $\micron$), and strong
molecular bands of FeH (1.194, 1.239 $\micron$) and H$_2$O (1.3 $\micron$) observed.
Above the spectrum we plot an opacity spectrum of FeH
\citep{dul03}, which shows that much of the fine
structure observed is caused by this molecule.
There are no lines of Mg I, Ca I, or Al I present in this spectral region,
and the $\phi$ TiO system around 1.25 $\micron$ \citep{jor94}
is either absent or obscured by FeH absorption.  The overall spectral
morphology at J-band is quite similar to the L7
DENIS 0205$-$1159AB,
with the notable exception of stronger FeH and H$_2$O absorption.

\subsection{Optical Spectrum}

We obtained a red optical spectrum of 2MASS 0532+8246 on 3 January 2003 (UT) using
LRIS on the Keck-I 10m Telescope.  Conditions were clear with
sub-arcsecond
seeing.  Two exposures of 1200 sec each were obtained through the red
channel, dithered by 2$\arcsec$ between exposures along the
1$\arcsec$ (4.7 pixels) slit.  We employed the 400 lines mm$^{-1}$
grating blazed at 8500 {\AA}, yielding 6300--10100 {\AA} spectra with
7 {\AA} resolution (R $\sim$ 1200).
Dispersion on the chip was 1.9 {\AA} pixel$^{-1}$.
The OG570 order-blocking filter was used to suppress higher-order light.
Observations of the B1 V flux standard
Hiltner 600 \citep{ham94}
were obtained for flux calibration, and the G0 V star HD 38847 was observed
after the target at similar airmass (2.4) for
telluric calibration.  HeNeAr arc lamp exposures were
taken immediately after the target observations for wavelength calibration,
and quartz lamp flat-field exposures (reflected off of the interior dome)
were observed at the start of the night to correct for detector response.
Data reduction procedures, using standard
IRAF\footnote{Image Reduction and Analysis Facility (IRAF) is distributed
by the National Optical Astronomy Observatories,
which are operated by the Association of Universities for Research
in Astronomy, Inc., under cooperative agreement with the National
Science Foundation.} routines,
are discussed in detail in \citet{kir03}.

The final flux-calibrated and telluric-corrected optical spectrum for 2MASS 0532+8246 is
included in Figure 2a and shown in detail in Figure 2c.  Strong metal hydride bands
of FeH (8692, 9896 {\AA}), CrH (8611, 9969 {\AA}), and CaH (6750 {\AA})
are clearly discernable.  The 9896 {\AA} FeH Wing-Ford
band is the strongest seen in any cool spectrum to date, and absorption from
the shorter wavelength FeH and CrH bands extends to roughly 9100 {\AA}.  Weaker bands
of TiO (7053, 8432 {\AA}) and H$_2$O (9250, 9400 {\AA}) are also seen.
We point out an absorption feature between 9570--9700 {\AA} which
is too strong to be attributable to
intrinsic H$_2$O absorption, based on the weakness of
the 9250 {\AA} band; it is also
not properly placed for FeH, CrH, TiH, or CH$_4$ absorption.  This band
is weakly present in late-type M and early-type L dwarf spectra, and may be a
hitherto unrecognized metal hydride band.  It is also possible, although unlikely,
that this feature is attributable to residual telluric absorption,
as telluric corrections were made only to 9650 {\AA} (the extent of the strong
terrestrial H$_2$O bands; Stevenson 1994).
Individual alkali lines of Na I (8183, 8195 {\AA}),  Rb I (7800, 7948 {\AA}),
and Cs I (8521, 8943 {\AA}) are present, while the resonance K I doublet
(7665, 7699 {\AA}) is exceptionally strong and pressure-broadened,
as seen in late-type L and T dwarfs \citep{kir99,lie00}.
The red optical/NIR colors of 2MASS 0532+8246 are clearly a
consequence of the K I red wing;
integrating the spectrum over the $I_C$ bandpass \citep{bes90} yields
$I_C$ $\approx$ 19.2 and $I_C - J$ $\approx$ 4.0, similar to mid- and late-type L dwarfs
\citep{dah02}.
The blue end of the spectrum appears to be similarly suppressed by
the pressure-broadened Na I doublet (5890, 5896 {\AA}).
Both Li I (6708 {\AA}) absorption and H$\alpha$ (6563 {\AA}) emission lines
are absent.
The overall spectral morphology, excluding hydride bands and TiO, is
again quite similar to the L7 DENIS 0205$-$1159AB, suggesting a rather cool
atmosphere.

\subsection{Characterization and Spectral Classification}

2MASS 0532+8246 is not a T dwarf, based on the absence of the
1.6 and 2.2 $\micron$ CH$_4$ bands \citep{me02a,geb02}, and the presence
of strong FeH and CrH and weak TiO and Na I features in the optical \citep{me03c}.
It appears that this object is instead a
metal-deficient, late-type L
dwarf.  Enhanced CIA H$_2$ and strong hydride bands
in the optical and NIR spectra of 2MASS 0532+8246 are consistent with
a metal-poor atmosphere.  Both are consequences of the overall reduced metal
opacity, resulting in larger column abundances of the remaining chemical
species and increased relative abundance of
metal hydrides over double-metal species.
The presence of the 7053 and 8432 {\AA} TiO bands is somewhat at odds
with this interpretation, particularly
since TiO is generally weak or absent in disk dwarfs later than
L5 \citep{kir99}.
One possibility is that the mechanism of Ti depletion in these objects -- incorporation into
multiple-metal condensates of
CaTiO$_3$, Ca$_3$Ti$_2$O$_7$, Ca$_4$Ti$_3$O$_{10}$, and Ti$_2$O$_3$ \citep{lod02} --
may be inhibited in the low-metallicity atmosphere of 2MASS 0532+8246.
Chemical equilibrium calculations
incorporating subsolar abundances are required to address this hypothesis.
The TiO bands may also indicate a somewhat warmer atmosphere than
the L7 comparison object DENIS 0205$-$1159AB.

Classifying 2MASS 0532+8246 is therefore not straightforward,
particularly as there are
no other metal-deficient L dwarfs currently known for direct comparison.
The M subdwarf and extreme subdwarf sequences of \citet{giz97} terminate at
sdM7 and esdM5.5, respectively, although recent late-type
discoveries are classified via
extrapolation and limited modifications of this scheme
\citep{lep03a,lep03b}.  Metal classification (d, sd, esd)
in the Gizis scheme is based on the strength of the
7053 {\AA} TiO band, which is exceedingly weak in the spectrum of 2MASS 0523+8246.
Numerical classification is based on the 6750 {\AA} CaH band, which is suppressed
in this spectrum by Na I and K I absorption.
Thus, a simple extrapolation of the Gizis scheme fails in the L dwarf regime.
While there is general
agreement between the spectral morphologies of 2MASS 0532+8246 and
the DENIS 0205$-$1159AB, the optical/J-band hydride features and
TiO bands are more similar to mid-type L dwarfs \citep{kir99,mcl03}.
Additionally, there
are no cool dwarf standards that match the 1.3--2.4 $\micron$ spectral region.
We therefore tentatively classify this object a late-type sdL.
Identification of additional metal-poor
L subdwarfs may eventually allow the definition of a robust classification scheme.

\section{Space Motion}

Examination of the IRCam and CCD followup images show that 2MASS 0532+8246 has
moved significantly since it was first imaged by
2MASS.  We therefore measured its proper motion by deriving its
coordinates on each of the images, using the
2MASS coordinates of background stars
for astrometric calibration.  Results are listed in Table 2.
A linear fit yields $\mu$ = 2$\farcs$60$\pm$0$\farcs$15 yr$^{-1}$ at
position angle 130.0$\pm$1.8$\degr$, making 2MASS 0532+8246 one of the
highest proper motion stars known \citep{bak02}.
This object is not present in the LHS because of its optical faintness, although
a faint counterpart may be present in the POSS-II R-band image
at the extrapolated position (Figure 1).
Residuals in the proper motion fit are indicative
of parallactic motion, but positional uncertainties are too large for a reliable measure.

The radial velocity of 2MASS 0532+8246 was derived by cross-correlating
its J-band spectrum with those of five early- and mid-type L dwarfs
(chosen for their strong NIR FeH absorption) with measured radial velocities:
2MASS 0746+2000AB, 2MASS 1439+1929, G 196-3B, Kelu 1, and DENIS 1228$-$1547AB
\citep{bas00,rei02}.
All five of these objects have been observed as part of
the NIRSPEC Brown Dwarf Spectroscopic Survey \citep{mcl03}, and are
therefore wavelength calibrated in the same manner as the data presented
here.  We derive a heliocentric $v_{rad} = -195{\pm}11$ km s$^{-1}$, where the
uncertainty is derived from the scatter of values amongst the comparison objects.
This value is
consistent with heliocentric velocity offsets in the 1.1690/1.1773 $\micron$ K I lines
(${\langle}v_{rad}{\rangle} = -175{\pm}17$ km s$^{-1}$) and the
optical Cs I and Rb I lines (${\langle}v_{rad}{\rangle} = -192{\pm}24$ km s$^{-1}$).

The substantial radial velocity and proper motion of
2MASS 0532+8246 identifies it as a high velocity source.
Determining its total space motion, however, requires an estimate
of its distance.  If we assume 2MASS 0532+8246 is bound to the Galaxy
(Galactocentric $v_{tot} \lesssim 500$ km s$^{-1}$; Carney, Latham, \& Laird 1988),
we derive an upper limit distance of 54 pc assuming
($U, V, W$)$_{\sun}$ = (10.00, 5.25, 7.17) km s$^{-1}$
\citep{deh98} and Local Standard of Rest (LSR) velocity
$V_{LSR} = 220$ km s$^{-1}$ \citep{ker86}.
A spectrophotometric parallax estimate is highly uncertain, given
the substantial redistribution of flux in the NIR and uncertainty
in the numeric subtype for this object.
However, assuming that the $M_{I_C}$ magnitude of
2MASS 0532+8246 is similar to that of mid- to late-type L dwarfs
\citep[$17 \lesssim M_{I_C} \lesssim 19$]{dah02}, we derive a
conservative estimate of 10--30 pc.  The corresponding heliocentric space
velocities at the median distance are
($U, V, W$) = ($-$13, $-$301, 28) km s$^{-1}$.  The substantial
V velocity, between $-$390 and $-$212 km s$^{-1}$ over the
adopted distance range, indicates zero or retrograde motion with respect to
the LSR, and lends
additional support that
this object is a member of the halo population.

\section{Is 2MASS 0532+8246 Substellar?}

Disk L dwarfs with spectral types later than $\sim$L4 are
predominately substellar \citep{giz00}; hence, the late-type spectral
morphology of 2MASS 0532+8246 suggests
that it too may be a brown dwarf.  To test this
hypothesis, we compared the estimated T$_{eff}$
of this object to subsolar metallicity evolutionary models from \citet{bur01}.
In Figure 3, we plot
T$_{eff}$ versus mass for ages 10 and 15 Gyr and metallicities
$Z$ = 0.1, 0.3, and 0.01 $Z_{\sun}$.  At these late ages, the transition between
stellar and substellar masses spans a substantial range in temperature, and is
more dramatic for lower metallicities due to the reduced atmospheric opacities
and hence enhanced luminosities \citep{bur93}.

Deriving an accurate T$_{eff}$ for 2MASS 0532+8246 is difficult
without adequate distance or bolometric flux measurements.
Solar-metallicity mid- to late-type L dwarfs have temperatures
$1400 \lesssim T_{eff} \lesssim 1800$ K
\citep{kir99,bas00,pav00,leg01,stp01,me02a,dah02}.  In general there is little
difference between the temperatures of M-type
subdwarfs and dwarf stars of the same numeric type
\citep{giz97,leg00}, although late-type M subdwarfs (e.g., the sdM7 LHS 377)
may be 100-300 K hotter \citep{leg00}.
We therefore adopt a conservative temperature range
of $1400 \lesssim T_{eff} \lesssim 2000$ K.  We further assume that
the metallicity of 2MASS 0532+8246 lies
in the range $0.01 \lesssim Z \lesssim 0.1$ $Z_{\sun}$,
typical for sdM and esdM stars \citep{giz97,giz97b}.
For these parameters, we derive a mass range of
0.077 $\lesssim$ M $\lesssim$ 0.085 M$_{\sun}$.
Larger masses correspond to lower metallicities,
so that in each case the derived mass is below the HBMM.
Therefore, we conclude the 2MASS 0532+8246 is probably substellar,
although more refined estimates of its atmospheric properties through
further observations and spectral
modelling are required to validate this result.

\section{Discussion}

The existence of a substellar subdwarf is not unexpected, as mass functions for
the nearby halo population \citep{giz99} and globular clusters \citep{pio99} are shallow
but rising: dN/dM $\propto$ M$^{-(0.5 - 1.3)}$
at M $\sim$ 0.15 M$_{\sun}$,
similar to the disk population \citep{rei99,cha02}.
However, the steep drop in T$_{eff}$, and hence luminosity, across the metal-poor
substellar boundary implies that the vast majority of these halo brown dwarfs will have
T$_{eff} \lesssim 1300$ K; i.e., in the T dwarf regime and cooler \citep{me02a}.
Indeed, one possible T-type subdwarf has already been identified, 2MASS 0937+2931
\citep{me02a}, an object which also exhibits unusually strong CIA H$_2$
and FeH absorption \citep{me03c} and
a substantial proper motion \citep{me03a}.
Relatively few L-type
subdwarfs should exist, however, as they span a much narrower range in mass at ages
later than 10 Gyr.   The discovery of 2MASS 0532+8246 may therefore be quite fortuitous.

Given its optical faintness, we expect that few counterparts to this
late-type L subdwarf will be identified in
current proper motion surveys.  NIR proper motion surveys, on the other
hand, could potentially reveal more of these objects by detecting them at the peak of their
spectral energy distribution.  A sizeable fraction of the sky ($\sim$30\%)
has been scanned more than once by 2MASS during its three years of
observations, and analysis of these data may identify
new, very cool high-motion stars.
A second generation 2MASS survey could
greatly improve on this
work, adding substantially to the short list
of very cool subdwarfs currently known.

\acknowledgments

A.\ J.\ B.\ acknowledges useful discussions with
K.\ Cruz and J.\ Wei-Chun during the
preparation of the manuscript,
and thanks our anonymous referee for her/his helpful criticisms.
We are grateful for the assistance of our
Keck Observing Assistants Joel Aycock and Gary Puniwai, and Instrument
Specialist Paola Amico;
and our Palomar Telescope Operators, Karl Dunscombe and Barrett ``Skip'' Staples.
We also acknowledge the 2MASS staff for their laudable efforts on the 2MASS database,
without which none of this research could have been possible.
A.\ J.\ B.\ acknowledges support by NASA
through Hubble Fellowship grant HST-HF-01137.01 awarded by the
Space Telescope Science Institute, which is operated by the
Association of Universities for Research in Astronomy, Inc., for
NASA, under contract NAS 5-26555.
A.\ S.\ B.\ acknowledges funding through NASA grants NAG5-10760 and NAG5-10629.
Portions of the data presented herein were obtained at the
W.\ M.\ Keck Observatory which is operated as a scientific partnership
among the California Institute of Technology, the University of California,
and the National Aeronautics and Space Administration. The Observatory was
made possible by the generous financial support of the W.\ M.\ Keck Foundation.
This publication makes use of data from the Two
Micron All Sky Survey, which is a joint project of the University
of Massachusetts and the Infrared Processing and Analysis Center,
funded by the National Aeronautics and Space Administration and
the National Science Foundation.
2MASS data were obtain through
the NASA/IPAC Infrared Science Archive, which is operated by the
Jet Propulsion Laboratory, California Institute of Technology,
under contract with the National Aeronautics and Space
Administration.
POSS-I and POSS-II images were obtained from the DSS
image server maintained by the Canadian Astronomy Data Centre,
which is operated by the Herzberg Institute of Astrophysics,
National Research Council of Canada.
The authors wish to extend special thanks to those of Hawaiian ancestry
on whose sacred mountain we are privileged to be guests.

\clearpage

\begin{deluxetable}{lc}
\tablecaption{Observational Properties of the Late-type sdL 2MASS 0532+8246.}
\tablewidth{0pt}
\tablehead{}
\startdata
${\alpha}_{J2000}$\tablenotemark{a} & 5$^h$ 32$^m$ 53$\farcs$46 \\
${\delta}_{J2000}$\tablenotemark{a} & +82$\degr$ 46$^m$ 46$\farcs$5 \\
$I_{C}$\tablenotemark{b} &  19.2 \\
Gunn $z$ & 18.30$\pm$0.16 \\
2MASS $J$ & 15.18$\pm$0.06 \\
2MASS $H$ & 14.90$\pm$0.10 \\
2MASS $K_s$ & 14.92$\pm$0.15 \\
$\mu$ & 2$\farcs$60$\pm$0$\farcs$15 yr$^{-1}$ \\
$\theta$ & 130.0$\pm$1.8$\degr$ \\
$v_{rad}$ & $-195{\pm}11$ km s$^{-1}$ \\
\enddata
\tablenotetext{a}{Epoch 1 March 1999 (UT).}
\tablenotetext{b}{Measured from optical spectrum; see $\S$ 3.2.}
\end{deluxetable}

\begin{deluxetable}{llcccc}
\tablecaption{Astrometry of 2MASS 0532+8246}
\tablewidth{0pt}
\tablehead{
\colhead{Epoch (UT)} &
\colhead{Instrument} &
\colhead{${\alpha}_{J2000}$} &
\colhead{${\delta}_{J2000}$} &
\colhead{${\sigma}_{\alpha}$\tablenotemark{a}} &
\colhead{${\sigma}_{\delta}$\tablenotemark{a}} \\
\colhead{(1)} &
\colhead{(2)} &
\colhead{(3)} &
\colhead{(4)} &
\colhead{(5)} &
\colhead{(6)} \\ }
\startdata
01 Mar 1999 02:20 & 2MASS & 5$^h$ 32$^m$ 53$\farcs$46 & +82$\degr$ 46$^m$ 46$\farcs$53 & 0$\farcs$30 & 0$\farcs$30 \\
25 Sep 1999 11:53 & P60 IRCAM & 5$^h$ 32$^m$ 54$\farcs$14 & +82$\degr$ 46$^m$ 45$\farcs$60 & 0$\farcs$07 & 0$\farcs$06  \\
18 Nov 2002 08:01 & P60 CCD & 5$^h$ 32$^m$ 57$\farcs$37 & +82$\degr$ 46$^m$ 40$\farcs$33 & 0$\farcs$35 & 0$\farcs$23 \\
25 Jan 2003 06:33 & P60 CCD & 5$^h$ 32$^m$ 57$\farcs$66 & +82$\degr$ 46$^m$ 40$\farcs$00 & 0$\farcs$37 & 0$\farcs$22 \\
\enddata
\tablenotetext{a}{Uncertainty in ($\alpha$, $\delta$) derived from standard deviation of
positions for background stars.}
\end{deluxetable}

\clearpage

\begin{figure}
\plotone{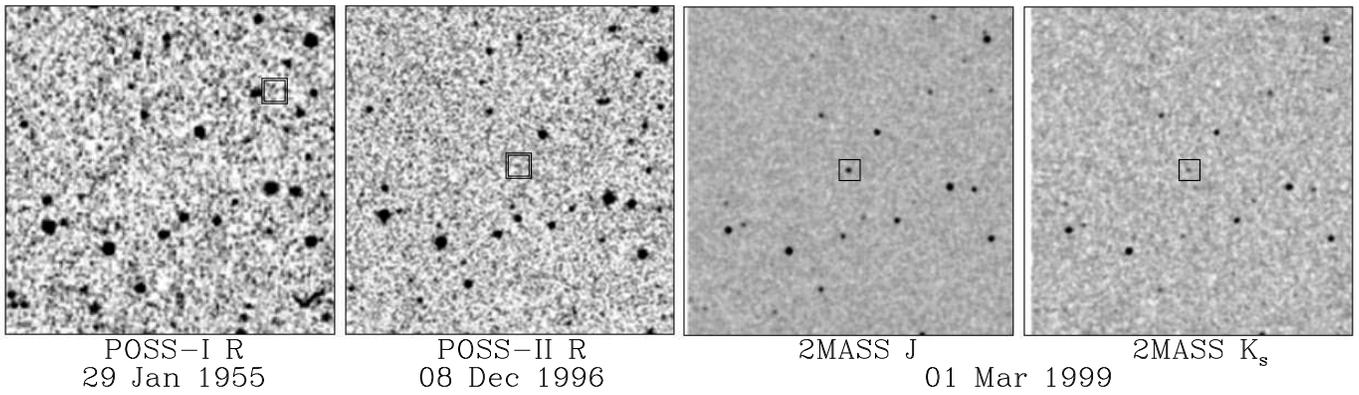}
\caption{Finderchart for 2MASS 0532+8246, showing POSS-I (R-band), POSS-II (R-band),
and 2MASS (J- and K$_s$-bands) fields.  Images are scaled to the same
spatial resolution, 5$\arcmin$ on a side, with North up and East
to the left.  A 10$\arcsec$ box is centered at the position of
the subdwarf in the 2MASS images and at the extrapolated position
(assuming $\mu = 2{\farcs}6$ and $\theta = 130\degr$) in the
POSS-I and POSS-II images.  \label{fig1} }
\end{figure}

\begin{figure}
\epsscale{0.65}
\plotone{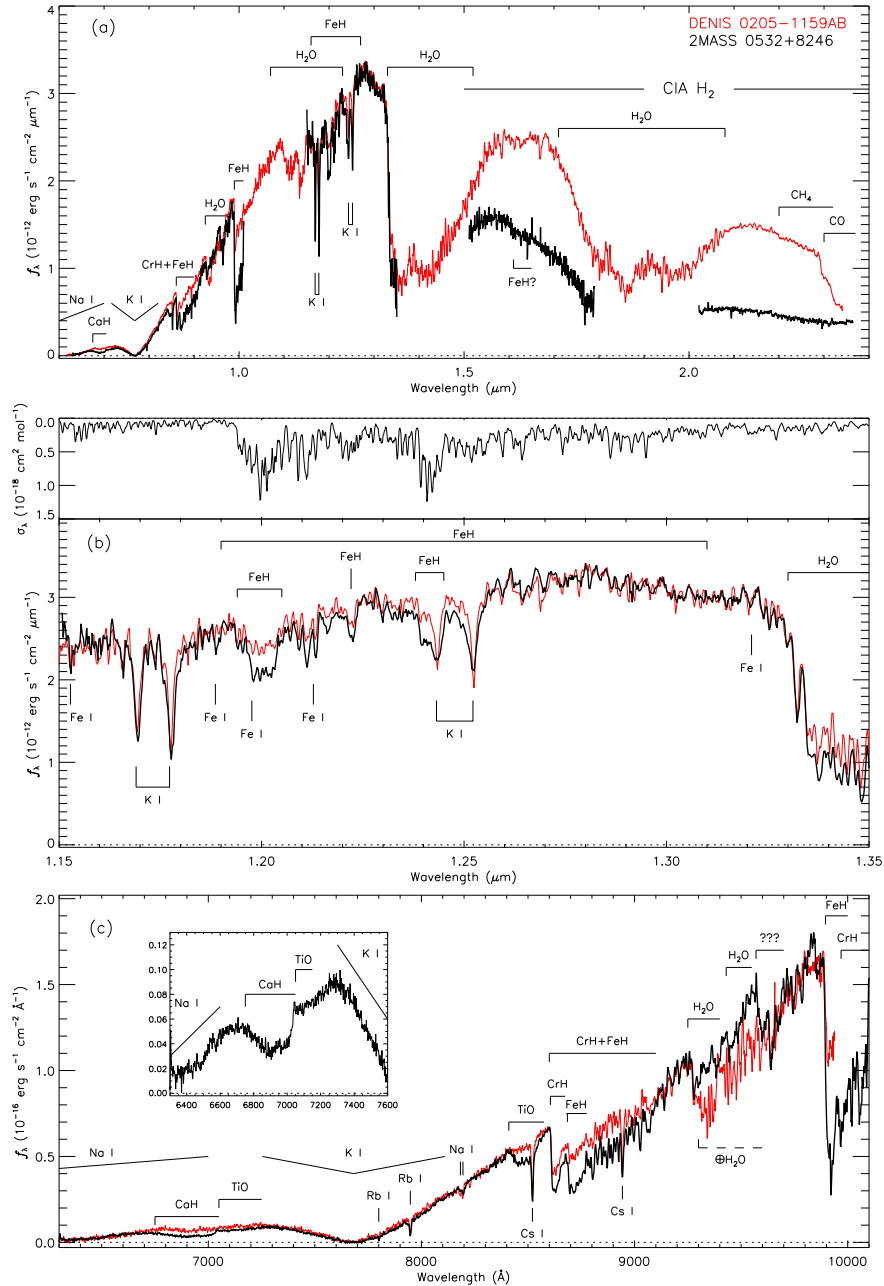}
\caption{The spectrum of 2MASS 0532+8246 (thick black line) as compared
to the L7 DENIS 0205$-$1159AB (thin grey line; data from Kirkpatrick et al.\ 1999;
McLean et al.\ 2003)  In all panels, spectral data for 2MASS 0532+8246 have been
shifted by $v_{rad} = +195$ km s$^{-1}$, and data for DENIS 0205$-$1159AB have been
scaled to coincide at 1.27 $\micron$.  Zeropoints are indicated by dotted lines.
(a) Observed 0.63--2.35 $\micron$ spectrum,
with NIRSPEC bands scaled to 2MASS photometry.  Key atomic and molecular features are
indicated; note that the 2.2 $\micron$ CH$_4$ and 2.3 $\micron$ CO bands
present in the spectrum of DENIS 0205$-$1159AB \citep{mcl01} are not present in that
of 2MASS 0532+8246.
(b) {\em Top:} FeH absorption coefficient versus wavelength, from \citet{dul03}. {\em Bottom:}
J-band spectrum of 2MASS 0532+8246,
with line identifications for K I, Fe I, FeH, and H$_2$O.
(c) Red optical spectrum, with key features indicated.
Uncorrected telluric H$_2$O absorption in the DENIS 0205$-$1159AB data
is indicated by the dashed bracket.
Inset window shows a close-up of the 6350--7600 {\AA} spectral region,
highlighting strong CaH and weak TiO bands; no Li I or H$\alpha$ lines
are seen.  \label{fig2} }
\end{figure}

\begin{figure}
\epsscale{0.9}
\plotone{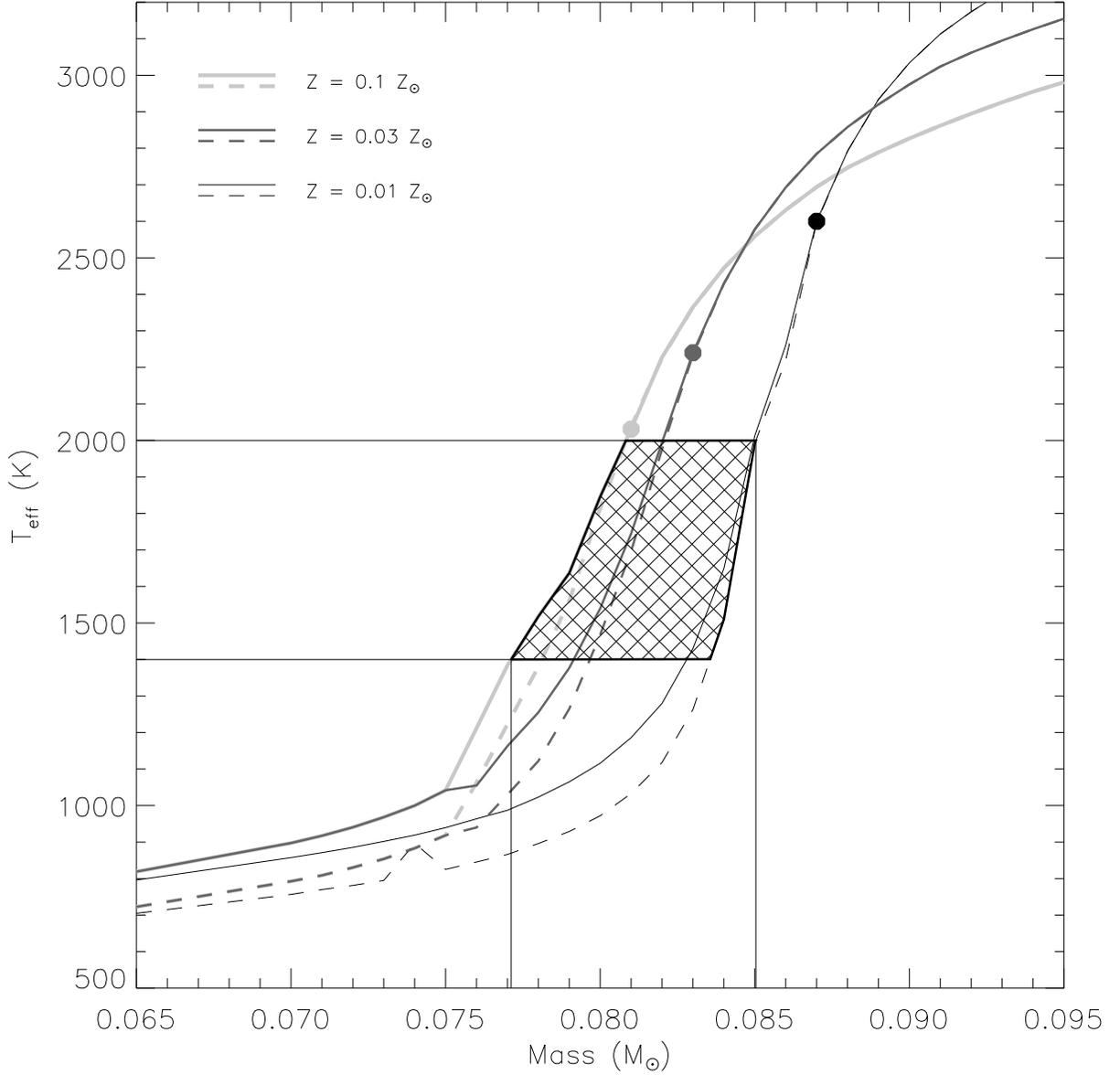}
\caption{T$_{eff}$ versus mass for theoretical evolutionary models with
$Z = 0.1$ (light grey), 0.03 (dark grey),
and 0.01 (black) $Z_{\sun}$, and ages 10 (solid) and 15 (dashed) Gyr.
HBMMs at 10 Gyr for all three metallicities are
indicated by solid circles.
Our adopted 1400 $\lesssim$ T$_{eff}$ $\lesssim$ 2000 K
range for 2MASS 0532+8246 corresponds
to masses of 0.077 $\lesssim$ M $\lesssim$ 0.085 M$_{\sun}$
over all values of metallicity and age shown (hatched region).  For each metallicity
the derived
mass is below the corresponding HBMM.
\label{fig3}}
\end{figure}


\begin{thebibliography}{}


\bibitem[Bakos, Sahu, \& N{\'{e}}meth(2002)]{bak02}Bakos, G.\ {\'{A}}.,
Sahu, K.\ C., \& N{\'{e}}meth, P. 2002, \apjs, 141, 187

\bibitem[Basri et al.(2000)]{bas00}Basri, G., Mohanty, S., Allard, F.,
Hauschildt, P.\ H., Delfosse, X., Mart{\'{\i}}n, E.\ L., Forveille, T.,
\& Goldman, B. 2000, \apj, 538, 363

\bibitem[Bessell(1990)]{bes90}Bessell, M.\ S. 1990, \pasp, 102, 1181

\bibitem[Borysow, J{\o}rgensen, \& Zheng(1997)]{bor97}Borysow, A., J{\o}rgensen,
U.\ G., \& Zheng, C. 1997, \aap, 324, 185

\bibitem[Burgasser, Kirkpatrick, \& Liebert(2003)]{me03c}Burgasser, A.\ J., Kirkpatrick, J.\ D.,
Liebert, J. 2003, \apj, submitted

\bibitem[Burgasser et al.(2003)]{me03a}Burgasser, A.\ J., Kirkpatrick,
J.\ D., McElwain, M.\ W., Cutri, R.\ M., Burgasser, A.\ J., \& Skrutskie, M.\ F.
2003, \aj, 125, 850

\bibitem[Burgasser et al.(2002)]{me02a} Burgasser, A.\ J., et al. 2002, \apj, 564, 421

\bibitem[Burrows et al.(2002)]{bur02} Burrows, A., Burgasser, A.\ J., Kirkpatrick, J.\ D.,
Liebert, J., Milsom, J.\ A., Sudarsky, D., \& Hubeny, I. 2002, \apj, 573, 394

\bibitem[Burrows et al.(2001)]{bur01}Burrows, A., Hubbard, W.\ B., Lunine,
J.\ I., \& Liebert, J. 2001, Rev.\ of Modern Physics, 73, 719

\bibitem[Burrows et al.(1993)]{bur93}Burrows, A., Hubbard, W.\ B., Saumon, D.,
\& Lunine, J.\ I. 1993, \apj, 406, 158


\bibitem[Carney, Latham, \& Laird(1988)]{car88}Carney, B., Latham, D., \& Laird,
J.\ B. 1988, \aj, 96, 560

\bibitem[Chabrier(2002)]{cha02}Chabrier, G. 2002, \apj, 567, 304

\bibitem[Chabrier \& Baraffe(1997)]{cha97}Chabrier, G., \& Baraffe, I.
1997, \aap, 327, 1039

\bibitem[Chamberlin \& Aller(1951)]{cha51}Chamberlin, J.\ W., \& Aller,
L.\ H. 1951, \apj, 114, 52

\bibitem[Cushing et al.(2003)]{cus03}Cushing, M.\ C., Rayner, J.\ T.,
Davis, S.\ P., \& Vacca, W.\ D. 2003, \apj, 582, 1066

\bibitem[Cutri et al.(2003)]{cut03}Cutri, R.\ M., et al. 2003,
\url{http://www.ipac.caltech.edu/2mass/releases/allsky/doc/explsup.html}

\bibitem[Dahn et al.(2002)]{dah02}Dahn, C.\ C., et al. 2002, \aj,
124, 1170

\bibitem[Dehnen \& Binney(1998)]{deh98}Dehnen, W., \& Binney, J.\ J. 1998, \mnras,
298, 387

\bibitem[Delfosse et al.(1997)]{del97}Delfosse, X., et al. 1997, \aap,
327, L25

\bibitem[Dulick et al.(2003)]{dul03} Dulick, M., Bauschlicher, C.\ W., Jr., Burrows, A.,
Sharp, C.\ M., Ram, R.\ S., \& Bernath, P. 2003, \apj, submitted

\bibitem[Epchtein et al.(1997)]{epc97}Epchtein, N., et al. 1997,
The Messenger, 87, 27

\bibitem[Fan et al.(2000)]{fan00} Fan, X., et al. 2000, \aj, 119, 928

\bibitem[Geballe et al.(2002)]{geb02}Geballe, T.\ R., et al. 2002, \apj, 564, 466

\bibitem[Gizis(1997)]{giz97}Gizis, J.\ E. 1997, \aj, 113, 806

\bibitem[Gizis et al.(2000)]{giz00}Gizis, J.\ E., Monet, D.\ G., Reid, I.\ N.,
Kirkpatrick, J.\ D., Liebert, J., \& Williams, R. 2000, \aj, 120, 1085

\bibitem[Gizis \& Reid(1997)]{giz97b}Gizis, J.\ E., \& Reid, I.\ N. 1997, \pasp, 109, 849

\bibitem[Gizis \& Reid(1999)]{giz99} ---. 1999,
\aj, 117, 508

\bibitem[Hamuy et al.(1994)]{ham94}Hamuy, M., Suntzeff, N.\ B., Heathcote, S.\ R., Walker, A.\ R.,
Gigoux, P., \& Phillips, M.\ M. 1994, PASP, 106, 566

\bibitem[J{\o}rgensen(1994)]{jor94}J{\o}rgensen, U.\ G. 1994, \aap, 284, 1791

\bibitem[Kent(1985)]{ken85}Kent, S.\ M. 1985, \pasp, 97, 165

\bibitem[Kerr \& Lynden-Bell(1986)]{ker86}Kerr, F.\ J., \& Lynden-Bell, D.
1986, \mnras, 221, 1023

\bibitem[Kirkpatrick et al.(1999)]{kir99}Kirkpatrick, J.\ D., et al. 1999,
\apj, 519, 802

\bibitem[Kirkpatrick et al.(2003)]{kir03} ---. 2003, \aj, in preparation

\bibitem[Kuiper(1939)]{kui39} Kuiper, G.\ P. 1939, \apj, 89, 549

\bibitem[Leggett et al.(2001)]{leg01} Leggett, S.\ K., Allard, F.,
Geballe, T., Hauschildt, P.\ H., \& Schweitzer, A. 2001, \apj, 548, 908

\bibitem[Leggett et al.(2000)]{leg00}Leggett, S.\ K., Allard, F., Dahn, C.,
Hauschildt, P.\ H., Kerr, T.\ H., \& Rayner, J. 2000, \apj, 535, 965

\bibitem[Lepine, Rich, \& Shara(2003)]{lep03a} Lepine, S.\
Rich, R.\ M., \& Shara, M.\ M. 2003, \aj, in press (astro-ph/0209284)

\bibitem[Lepine, Shara, \& Rich(2002)]{lep02} Lepine, S.\ Shara, M.\ M., \&
Rich, R.\ M. 2002, \aj, 124, 1190

\bibitem[Lepine, Shara, \& Rich(2003)]{lep03b} Lepine, S.\ Shara, M.\ M., \&
Rich, R.\ M. 2003, \apj, 585, L69

\bibitem[Liebert \& Probst(1987)]{lie87}Liebert, J., \& Probst, R.\ G. 1987, \araa, 25, 473

\bibitem[Liebert et al.(2000)]{lie00}Liebert, J., Reid, I.\ N., Burrows, A.,
Burgasser, A.\ J., Kirkpatrick, J.\ D., \& Gizis, J.\ E. 2000, \apj, 533,
L155

\bibitem[Lodders(2002)]{lod02}Lodders, K. 2002, \apj, 577, 974

\bibitem[Luyten(1979a)]{luy79a}Luyten, W.\ J. 1979a, LHS Catalogue: A Catalogue of Stars with Proper
Motions Exceeding 0$\farcs$5 Annually (Minneapolis: Univ.\ Minn.\ Press)

\bibitem[Luyten(1979b)]{luy79b} ---. 1979b, New Luyten Catalogue of
Stars with Proper Motions Larger than Two Tenths
of an Arcsecond (NLTT) (Minneapolis: Univ.\ Minn.\ Press)

\bibitem[Mart{\'{\i}}n et al.(1999)]{mrt99}Mart{\'{\i}}n, E.\ L., Delfosee, X.,
Basri, G., Goldman, B., Forveille, T., \& Zapatero Osorio, M.\ R. 1999, \aj, 118, 2466

\bibitem[McLean et al.(2000)]{mcl00}McLean, I.\ S., Graham, J.\ R.,
Becklin, E.\ E., Figer, D.\ F., Larkin, J.\ E., Levenson, N.\ A., \&
Teplitz, H.\ I. 2000, SPIE, 4008, 1048

\bibitem[McLean et al.(2003)]{mcl03}McLean, I.\ S., McGovern, M.\ R., Burgasser, A.\ J.,
Kirkpatrick, J.\ D., Prato, L., \& Kim, S. 2003, \apj, submitted

\bibitem[McLean et al.(2001)]{mcl01}McLean, I.\ S., Prato, L., Kim, S.\ S.,
Wilcox, M.\ K., Kirkpatrick, J.\ D., \& Burgasser, A.\ J. 2001, \apj, 561, L115

\bibitem[McLean et al.(1998)]{mcl98}McLean, I.\ S., et al. 1998, SPIE, 3354, 566

\bibitem[Mould \& Hyland(1976)]{mou76}Mould, J.\ R., \& Hyland, A.\ R. 1976, \apj, 208, 399

\bibitem[Murphy et al.(1995)]{mur95}Murphy, D.\ C., Persson, S.\ E.,
Pahre, M.\ A., Sivaramakrishnan, A., \& Djorgovski, S. G. 1995, \pasp, 107, 1234

\bibitem[Oke et al.(1995)]{oke95}Oke, J.\ B., et al. 1995, \pasp, 107, 375

\bibitem[Pavlenko, Zapatero Osorio, \& Rebolo(2000)]{pav00}Pavlenko, Ya.,
Zapatero Osorio, M.\ R., \& Rebolo, R. 2000, \aap, 355, 245

\bibitem[Piotto \& Zoccali(1999)]{pio99}Piotto, G., \& Zoccali, M. 1999, \aap, 345, 485

\bibitem[Reid \& Hawley(2000)]{rei00b}Reid, I.\ N., \& Hawley, S.\ L.
2000, New Light on Dark Stars (Chichester: Praxis)


\bibitem[Reid et al.(2002)]{rei02}Reid, I.\ N., Kirkpatrick, J.\ D.,
Liebert, J., Gizis, J.\ E., Dahn, C.\ C., \&
Monet, D.\ G. 2002, \aj, 124, 519

\bibitem[Reid et al.(1999)]{rei99}Reid, I.\ N., et al.  1999, \apj, 521, 613

\bibitem[Sandage \& Eggen(1959)]{san59}Sandage, A.\ R., \& Eggen, O.\ J.
1958, \mnras, 119, 278

\bibitem[Saumon et al.(1994)]{sau94}Saumon, D., Bergeron, P., Lunine, J.\ I.,
Hubbard, W.\ B., \& Burrows, A. 1994, \apj, 424, 333

\bibitem[Scholtz et al.(2000)]{shz00} Scholtz, R.-D., Irwin, M., Ibata, R.,
Jahrei$\beta$, H., \& Malkov, O.\ Yu. 2000, \aap, 353, 958

\bibitem[Schweitzer et al.(1999)]{sch99}Schweitzer, A., Shultz, R.-D., Stauffer, J.,
Irwin, M., \& McCaughren, M.\ J. 1999, \aap, 350, L62

\bibitem[Skrutskie et al.(1997)]{skr97}Skrutskie, M.\ F., et al. 1997, in The Impact of Large-Scale Near-IR
Sky Surveys, ed. F.\ Garzon (Dordrecht: Kluwer), p.\ 25

\bibitem[Smith et al.(2002)]{smi02}Smith, J.\ A., et al. 2002, \aj, 123, 2121

\bibitem[Stephens et al.(2001)]{stp01}Stephens, D.\ C., Marley, M.\ S.,
Noll, K.\ S., \&  Chanover, N. 2001, \apj, 556, L97

\bibitem[Stevenson(1994)]{ste94}Stevenson, C.\ C. 1994, \mnras, 267, 904

\bibitem[Tokunaga \& Kobayashi(1999)]{tok99}Tokunaga, A.\ T., \&
Kobayashi, N. 1999, \aj, 117, 1010

\bibitem[York et al.(2000)]{yor00} York, D.\ G., et al. 2000, AJ,
120, 1579

\end{thebibliography}
\end{document}